\documentclass[a4paper,11pt]{article}
\pdfoutput=1 

\usepackage{jcappub} 

\usepackage[T1]{fontenc} 

\newcommand{\orcid}[1]{\href{https://orcid.org/#1}{\,\includegraphics[width=8px]{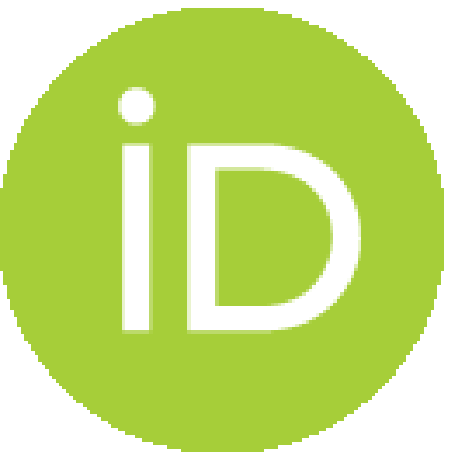}}}

\title{\boldmath Minimal model dependent constraints on cosmological nuisance parameters and cosmic curvature from combinations of cosmological data}

\author[a,1]{Bikash R. Dinda\orcid{0000-0001-5432-667X},\note{Corresponding author.}}

\affiliation[a]{Department of Physical Sciences, Indian Institute of Science Education and Research Kolkata, Mohanpur, Nadia, West Bengal 741246, India}

\emailAdd{bikashdinda.pdf@iiserkol.ac.in}

\abstract{The study of cosmic expansion history and the late time cosmic acceleration from observational data depends on the nuisance parameters associated with the data. For example, the absolute peak magnitude of type Ia supernova associated with the type Ia supernova observations and the comoving sound horizon at the baryon drag epoch associated with baryon acoustic oscillation observations are two nuisance parameters. The nuisance parameters associated with the the gamma-ray bursts data are also considered. These nuisance parameters are constrained by combining the cosmological observations using the Gaussian process regression method with minimal model dependence. The bounds obtained in this method can be used as the prior for the data analysis while considering the observational data accordingly. Along with these nuisance parameters, the cosmic curvature density parameter is also constrained simultaneously. We find that the constraints on the cosmic curvature density parameter show no significant deviations from a flat Universe.}

\begin{document}
\maketitle
\flushbottom

\section{Introduction}

The late time cosmic acceleration is one of the major discoveries in the late 20th-century \cite{SupernovaCosmologyProject:1997zqe,SupernovaSearchTeam:1998fmf,SupernovaCosmologyProject:1998vns,2011NatPh...7Q.833W}. This acceleration has been confirmed by several cosmological observations such as type Ia supernova \cite{Linden2009CosmologicalPE,Camarena:2019rmj,Pan-STARRS1:2017jku,Camlibel:2020xbn}, cosmic microwave background (CMB) \cite{Planck:2013pxb,Planck:2015fie,Planck:2018vyg}, baryon acoustic oscillation (BAO) \cite{BOSS:2016wmc,eBOSS:2020yzd,Hou:2020rse} etc. One of the possible explanations of the late time cosmic acceleration is the introduction of an exotic matter component, called dark energy, that has large negative pressure \cite{Copeland:2006wr,Peebles:2002gy,Yoo:2012ug}. Another alternative to explain this acceleration is the modification of general relativity at large cosmological scales \cite{Clifton:2011jh,Koyama:2015vza,Tsujikawa:2010zza,Joyce:2016vqv}. Several dark energy and modified gravity models have been proposed in the literature and these models have been constrained and shaped by the cosmological observations \cite{Joyce:2016vqv,Lonappan:2017lzt}.

Alongside the type Ia supernova, CMB, and BAO observations, several other observations have been continuously developed to better understand the nature of the late time cosmic acceleration and the mechanisms to explain this acceleration. For example, such observations are the cosmic chronometers (CC) \cite{Jimenez:2001gg,Pinho:2018unz} and gamma-ray bursts (GRB) \cite{Khadka:2021vqa,Wei:2008kq,Amati:2008hq} observations. All these recent observations are providing us with percent precision data which are helpful to understand the nature of dark energy or modification of gravity \cite{Weinberg_2013,Moresco:2022phi} with high precision.

The simplest candidate for dark energy is considered to be the cosmological constant and the corresponding model is called the $\Lambda$CDM model \cite{Carroll:2000fy}. This is the most successful model to study the evolution of the Universe in light of recent cosmological observations \cite{Planck:2013pxb,Planck:2015fie,Planck:2018vyg}. This model has theoretical problems like fine-tuning and cosmic coincidence problems \cite{Zlatev:1998tr,Sahni:1999gb,Velten:2014nra,Malquarti:2003hn}. Despite these theoretical problems, there are other inconsistencies with this model that arose from the percent precision observational data \cite{Perivolaropoulos:2021jda,Bull:2015stt}. One such example is the so-called Hubble tension \cite{DiValentino:2021izs,Krishnan:2021dyb,Vagnozzi:2019ezj}. Considering these caveats into account, it is necessary to go beyond the $\Lambda$CDM model. In this context, in literature, many dynamical dark energy and modified gravity models have been developed gradually \cite{Joyce:2016vqv,Lonappan:2017lzt}.

The background dynamics of the Universe are not only studied through different dark energy and modified gravity models but it has also been studied through different parametrizations to the different cosmological quantities without assuming any model \cite{Chevallier:2000qy,Linder:2002et,Barboza:2008rh,Thakur:2012rp,Dinda:2021ffa,Dinda:2019mev}. Even in recent times, background dynamics are studied from several observations with the minimal model or parametrization dependence by using various modern techniques \cite{Tutusaus:2018ulu,Zheng:2020tau,LHuillier:2018rsv,Wang:2019yob,Liu:2020pfa,Escamilla-Rivera:2019hqt}. One such technique is the Gaussian process regression (GPR) and it has been used in the literature in several contexts \cite{Seikel_2012,Shafieloo_2012,Hwang:2022hla}. The GPR method is the key ingredient in our analysis.

Most of the above-mentioned observations involve cosmological nuisance parameters like the peak absolute magnitude of the type Ia supernova \cite{Camarena:2021jlr,Gomez-Valent:2021hda}. Because of this reason, both the model dependent and independent studies of the late time cosmic acceleration depend on these cosmological nuisance parameters \cite{Cao:2022ugh,Colgain:2022nlb}. So, to study the late time cosmic acceleration, we need prior information on these parameters. However, without priors or with flat priors, these parameters can be constrained by combining different observations \cite{Dinda:2022jih}. In our analysis, we constrain these nuisance parameters from the joint analysis of cosmological observations, mentioned above.

Constraints on the nuisance parameters degenerate to the cosmic curvature density parameter \citep{Liu:2020pfa,Cai:2015pia}. Thus, it is also important to simultaneously constrain the cosmic curvature density parameter. For this reason, in our analysis, we keep the cosmic curvature density parameter as a free parameter and simultaneously constrain it with other parameters from different combinations of cosmological observations.

This paper is organized as follows: In Sec.~\ref{sec-basic}, we mention the basic cosmological equations; in Sec.~\ref{sec-methods}, we present the methodology to obtain bounds on several nuisance parameters from the combinations of different data; in Sec.~\ref{sec-results}, we present the results of our analysis; the concluding remarks are mentioned in Sec.~\ref{sec-conclusion}.

\section{Basics}
\label{sec-basic}

In our entire analysis, we assume that the geometry of the Universe is described by the Friedmann-Lema\^itre-Robertson-Walker (FLRW) metric. In FLRW cosmology, the line of sight comoving distance, $d_C$ is given as

\begin{equation}
d_C = c d_P,
\label{eq:loscov}
\end{equation}

\noindent
where $c$ is the speed of light in vacuum and $d_P$ is given as

\begin{equation}
d_P = \int_0^z \frac{d\tilde{z}}{ H(\tilde{z}) },
\label{eq:H_to_dP}
\end{equation}

\noindent
where $H$ is the Hubble parameter, $z$ is the redshift, and $\tilde{z}$ is the dummy argument for the redshift.

In the presence of the cosmic curvature, the transverse comoving distance differs from the line of sight comoving distance. The transverse comoving distance, $d_M$ is given as

\begin{equation}
    d_M = \begin{cases}
    \frac{c}{\sqrt{W_{\rm k0}}} \sinh \left( \sqrt{W_{\rm k0}} d_P \right), & \mbox{for } \Omega_{\rm k0}>0, \\
    c d_P, & \mbox{for } \Omega_{k0} = 0, \\
    \frac{c}{\sqrt{|W_{\rm k0}|}} \sin \left( \sqrt{|W_{\rm k0}|} d_P \right), & \mbox{for } \Omega_{\rm k0}<0, \\
    \end{cases}
    \label{eq:trnscov}
\end{equation}

\noindent
where $\Omega_{\rm k0}$ is the cosmic curvature density parameter given as $\Omega_{\rm k0}=-K c^2/a_0^2 H_0^2$; $K$ is the spatial curvature of the space-time; $a_{0}$ is the present value of the scale factor and it can be normalized as $a_{0}=1$ without loss of generalization; $H_0$ is the Hubble parameter at present or Hubble constant; $W_{\rm k0}$ is given as

\begin{equation}
W_{\rm k0} = \Omega_{\rm k0} H_0^2 = 10^4 \left( \Omega_{\rm k0} h^2 \right) \left[ {\rm Km}{\rm S}^{-1}{\rm Mpc}^{-1} \right]^2,
\label{eq:defnWk0}
\end{equation}

\noindent
where Hubble constant is related to the dimensionless parameter, $h$ given as $H_0=100 h {\rm Km}{\rm S}^{-1}{\rm Mpc}^{-1}$. The luminosity distance, $d_L$ is related to the transverse comoving distance as

\begin{equation}
d_L = (1+z) d_M.
\label{eq:dM_to_dL}
\end{equation}

\noindent
The angular diameter distance, $d_A$ is also related to the transverse comoving distance given as

\begin{equation}
d_A = \frac{d_M}{1+z}.
\label{eq:dM_to_dA}
\end{equation}

For a type Ia supernova, the observed (apparent) peak magnitude, $m$ is related to the luminosity distance given as

\begin{equation}
m - M_B = 5 \log_{10}{ \left( \frac{d_L}{ \text{Mpc} } \right) } + 25,
\label{eq:dL_to_m_basic}
\end{equation}

\noindent
where $M_B$ is the absolute peak magnitude of the type Ia supernova.

\subsection{Hubble parameter from luminosity distance}

In the subsequent sections, we will need the derivation of the Hubble parameter from the luminosity distance, because we will avoid any integral equation like in Eq.~\eqref{eq:H_to_dP}. To do this, we first take the derivative of Eq.~\eqref{eq:trnscov} and we get

\begin{equation}
    d'_M = \begin{cases}
    c d'_P \cosh \left( \sqrt{W_{\rm k0}} d_P \right), & \mbox{for } \Omega_{\rm k0}>0, \\
    c d'_P, & \mbox{for } \Omega_{k0} = 0, \\
    c d'_P \cos \left( \sqrt{-W_{\rm k0}} d_P \right), & \mbox{for } \Omega_{\rm k0}<0, \\
    \end{cases}
    \label{eq:trnscov_derivative}
\end{equation}

\noindent
where prime denotes the derivative with respect to the redshift. The above equation is rewritten as

\begin{equation}
    d'_P = \begin{cases}
    \frac{d'_M}{c \cosh \left( \sqrt{W_{\rm k0}} d_P \right)}, & \mbox{for } \Omega_{\rm k0}>0, \\
    \frac{d'_M}{c}, & \mbox{for } \Omega_{k0} = 0, \\
    \frac{d'_M}{c \cos \left( \sqrt{-W_{\rm k0}} d_P \right)}, & \mbox{for } \Omega_{\rm k0}<0. \\
    \end{cases}
    \label{eq:dP_derivative_trnscv}
\end{equation}

\noindent
We solve Eq.~\eqref{eq:trnscov} to get $d_P$ from $d_M$  and putting these solutions in the above equation, we get

\begin{equation}
    (d'_P)^2 = \begin{cases}
    \frac{(d'_M)^2}{c^2+W_{\rm k0}d_M^2}, & \mbox{for } \Omega_{\rm k0}>0, \\
    \frac{(d'_M)^2}{c^2}, & \mbox{for } \Omega_{k0} = 0, \\
    \frac{(d'_M)^2}{c^2+W_{\rm k0}d_M^2}, & \mbox{for } \Omega_{\rm k0}<0. \\
    \end{cases}
    \label{eq:dP_derivative_trnscov_alone}
\end{equation}

\noindent
The above equation can be rewritten as

\begin{equation}
(d'_P)^2 = \frac{(d'_M)^2}{c^2+W_{\rm k0}d_M^2},
\label{eq:dP_derivative_final}
\end{equation}

\noindent
and valid for all values of $\Omega_{\rm k0}$. Taking derivative of Eq.~\eqref{eq:H_to_dP}, we get $d'_P=1/H$. Putting this in the above equation we get

\begin{equation}
H^2 = \frac{c^2+W_{\rm k0}d_M^2}{(d'_M)^2}.
\label{eq:Hsqr_wrt_dM_dMp}
\end{equation}

\noindent
Using Eq.~\eqref{eq:dM_to_dL} in the above equation, we get

\begin{equation}
H^2 = \frac{(1+z)^2\left[c^2(1+z)^2+W_{\rm k0}d_L^2\right]}{\left[(1+z)d'_L-d_L\right]^2}.
\label{eq:Hsqr_wrt_dL_dLp}
\end{equation}

\noindent
Consequently, we get the Hubble parameter from the luminosity distance and its derivative given as

\begin{equation}
H = \left( \frac{(1+z)^2\left[c^2(1+z)^2+W_{\rm k0}d_L^2\right]}{\left[(1+z)d'_L-d_L\right]^2} \right)^{\frac{1}{2}}.
\label{eq:Hubble}
\end{equation}

\subsection{$d_L$ and $d'_L$ from $m$ and $m'$}

The luminosity distance can be computed from $m$ using Eq.~\eqref{eq:dL_to_m_basic} given as

\begin{equation}
d_L = 10^{\frac{1}{5} \left( m-25-M_B \right) } \hspace{0.2 cm} \text{Mpc}.
\label{eq:dL_from_m}
\end{equation}

Taking derivative of the above equation w.r.t redshift, we get $d'_L$ from $m$ and $m'$ given as

\begin{equation}
d'_L = \frac{\log{10}}{5} 10^{\frac{m-M_B-25}{5}} m' \hspace{0.2 cm} \text{Mpc},
\label{eq:dLp_from_mp}
\end{equation}

\section{Methodology}
\label{sec-methods}

We combine different cosmological observations to obtain bounds on different cosmological nuisance parameters. To do this, we consider the Pantheon sample for type Ia supernova observations \cite{Pan-STARRS1:2017jku}, the cosmic chronometers observations \citep{Jimenez:2001gg,Pinho:2018unz}, the baryon acoustic oscillation observations \citep{eBOSS:2020yzd}, and the Amati correlated gamma-ray bursts data \cite{Khadka:2021vqa,Wei:2008kq,Amati:2008hq}. For the methodology, here, we adapt the Gaussian process regression (GPR) analysis \citep{williams1995gaussian,GpRasWil,Seikel_2012,Shafieloo_2012,Hwang:2022hla,Keeley:2020aym}. Throughout this paper, we denote type Ia supernova, cosmic chronometers, baryon acoustic oscillation, and gamma-ray bursts data by 'SN', 'CC', 'BAO', and 'GRB' respectively. Any quantity having subscript or superscript with these keywords corresponds to the values of that quantity at data points of corresponding observations. For example, $z_{\rm SN}$ represents the redshift points corresponding to the SN data.

\subsection{Reconstruction of $m$ and $m'$ from SN data using GPR}
\label{sec-SN_GPR}

This work aims to combine the SN data with other data, for example, the CC data. Comparing any two data sets are not straightforward because two different observations have data for two different quantities and they are not at the same data points. For example, the SN data measures $m$ (with error bars, $\Delta m$) \cite{Pan-STARRS1:2017jku} and the CC data measures $H$ (with error bars, $\Delta H$) \citep{Jimenez:2001gg,Pinho:2018unz} at different redshift points. We consider the notation $\Delta X$ to represent the standard deviation in any quantity, $X$ throughout this analysis.

For the solution of this problem, GPR is useful \citep{williams1995gaussian,GpRasWil,Seikel_2012,Shafieloo_2012,Hwang:2022hla,Keeley:2020aym}. To have the lower computational time cost, we choose the posterior GPR approach \citep{williams1995gaussian,GpRasWil,Seikel_2012,Shafieloo_2012}. With this GPR analysis, we reconstruct a function for $m(z)$ and the corresponding uncertainties from the SN data. The posterior approach of the GPR analysis takes the data and the uncertainties in the data as the inputs and predicts the function for the corresponding quantity and the associated uncertainty in it. For these predictions, GPR analysis also considers a kernel covariance function and a mean function as inputs. For this purpose, we choose the squared exponential kernel and the mean function for $m(z)$ corresponding to the $\Lambda$CDM model. For details of the GPR analysis, see Appendix~\ref{sec-gpr_basic} and for details of the kernel and mean functions, see Appendix~\ref{sec-parameters_dependence}. We denote the predicted function for $m$ as $m_{\rm GPR}$ and the corresponding uncertainties as $\Delta m_{\rm GPR}$.

The GPR analysis not only predicts the function for a quantity corresponding to the observational data, but it can also predict the function for the derivatives of the quantity. We only need the first derivative prediction in our analysis. For the details of the first derivative prediction, see Appendix~\ref{sec-gpr_basic}. We denote the predicted function for the first derivative of $m$ w.r.t $z$ as $m'_{\rm GPR}$ and the corresponding uncertainties as $\Delta m'_{\rm GPR}$. All these predicted functions will be used in the next steps.

\subsection{Combination of SN and CC data: bounds on $M_B$ and $\Omega_{\rm k0}h^2$}
\label{sec-CCvsSN}

From reconstructed $m_{\rm GPR}$ (obtained from the above subsection), we reconstruct $d_L$ using Eq.~\eqref{eq:dL_from_m} and the corresponding uncertainties. Similarly, from reconstructed $m_{\rm GPR}$ and $m'_{\rm GPR}$ (obtained from the above subsection), we reconstruct $d'_L$ using Eq.~\eqref{eq:dLp_from_mp} and the corresponding uncertainties. Note that, these reconstructed functions are dependent on the parameter, $M_B$. This can be seen through Eqs.~\eqref{eq:dL_from_m} and~\eqref{eq:dLp_from_mp}.

From these reconstructed functions, next, we reconstruct the corresponding Hubble parameter using Eq.~\eqref{eq:Hubble}. From Eq.~\eqref{eq:Hubble}, we can see that reconstruction of $H$ from $d_L$ and $d'_L$ depends on the $W_{\rm k0}$ parameter. So, ultimately, the reconstructed $H$ and the corresponding uncertainties are dependent on both $M_B$ and $W_{\rm k0}$. We denote these as $H_{\rm GPR}$ and $\Delta H_{\rm GPR}$ respectively.

We compare the reconstructed Hubble parameter with the Hubble parameter data obtained directly from the CC data to determine $M_B$ and $W_{\rm k0}$ simultaneously. For this purpose, we define a log-likelihood ($\log{L}_{\rm CC}$) given as

\begin{eqnarray}
\log{L}_{\rm CC} (M_B,W_{\rm k0}) &=& - \frac{1}{2} \sum_{z_{\rm CC}} \frac{ \left[ H_{\rm GPR}(z_{\rm CC},M_B,W_{\rm k0})-H_{\rm CC} \right]^2}{ \Delta H_{\rm tot}^2(z_{\rm CC},M_B,W_{\rm k0}) } \nonumber\\
&& - \frac{1}{2} \sum_{z_{\rm CC}} \log{ \left[ 2 \pi \Delta H_{\rm tot}^2(z_{\rm CC},M_B,W_{\rm k0}) \right] },
\label{eq:lnlk_SN_CC}
\end{eqnarray}

\noindent
where $\Delta H_{\rm tot}^2$ is the total variance in the Hubble parameter given as $\Delta H_{\rm tot}^2 (z_{\rm CC},M_B,W_{\rm k0}) = \Delta H_{\rm GPR}^2 (z_{\rm CC},M_B,W_{\rm k0}) + \Delta H_{\rm CC}^2$. $H_{\rm CC}$ and $\Delta H_{\rm CC}$ correspond to the Hubble parameter data and the corresponding uncertainties obtained directly from the CC data. $z_{\rm CC}$ corresponds to the CC redshift points. We get simultaneous constraints on $M_B$ and $W_{\rm k0}$ by minimizing the negative log-likelihood mentioned in Eq.~\eqref{eq:lnlk_SN_CC}. Since, $W_{\rm k0}$ and $\Omega_{\rm k0}h^2$ are equivalent through Eq.~\eqref{eq:defnWk0}, finally, we get simulataneous constraints on $M_B$ and $\Omega_{\rm k0}h^2$.

\subsection{Combination of SN, CC and BAO data: constraints on $M_B$, $\Omega_{k0}h^2$, and $r_d$ (comoving sound horizon at baryon drag epoch)}
\label{sec-SN_BAO}

The BAO data have two kinds of measurements: one is in the line of sight direction and related to the quantity, $D_H$; the second one is in the transverse direction and related to the quantity, $D_M$. $D_H$ and $D_M$ are given as

\begin{eqnarray}
D_H &=& \frac{1}{r_d} \frac{c}{H}, 
\label{eq:DH_from_H} \\
D_M &=& \frac{1}{r_d} \frac{d_L}{1+z},
\label{eq:DM_from_dL}
\end{eqnarray}

\noindent
respectively, where $r_d$ is the comoving sound horizon at baryon drag epoch \cite{eBOSS:2020yzd}. We denote the BAO data which are related to $D_H$ as BAO1 and we denote the other BAO data which are related to $D_M$ as BAO2. So, BAO data can be considered as two types of data as BAO1 and BAO2 i.e. BAO=BAO1+BAO2.

Similar to the previous case, we reconstruct $H$ at BAO redshift points ($z_{\rm BAO}$) from the reconstructed $m_{\rm GPR}$ and $m'_{\rm GPR}$ (obtained from Sec.~\ref{sec-SN_GPR}) via $d_L$ and $d'_L$ and the corresponding uncertainties. Note that the BAO1 and BAO2 redshift points are the same and we call these as BAO redshift points, $z_{\rm BAO}$ i.e. $z_{\rm BAO1}=z_{\rm BAO2}=z_{\rm BAO}$. From this reconstructed Hubble parameter, we compute $D_H$ and the corresponding uncertainties at BAO redshift points using Eq.~\eqref{eq:DH_from_H}. We denote these as $D_H^{\rm GPR}$ and $\Delta D_H^{\rm GPR}$ respectively. We compare these with the $D_H$ directly obtained from the BAO1 data and we denote this as $D_H^{\rm BAO1}$. We denote the corresponding uncertainties as $\Delta D_H^{\rm BAO1}$. The corresponding log likelihood ($\log{L}_{\rm BAO1}$) is defined as

\begin{eqnarray}
\log{L}_{\rm BAO1} (M_B,W_{\rm k0},r_d) &=& - \frac{1}{2} \sum_{z_{\rm BAO}} \frac{ \left[ D_H^{\rm GPR}(z_{\rm BAO},M_B,W_{\rm k0},r_d)-D_H^{\rm BAO1} \right]^2}{ \left(\Delta D_H^{\rm tot}(z_{\rm BAO},M_B,W_{\rm k0},r_d)\right)^2 } \nonumber\\
&& - \frac{1}{2} \sum_{z_{\rm BAO}} \log{ \left[ 2 \pi \left(\Delta D_H^{\rm tot}(z_{\rm BAO},M_B,W_{\rm k0},r_d)\right)^2 \right] },
\label{eq:lnlk_CC_BAO_1}
\end{eqnarray}

\noindent
where $\left(\Delta D_H^{\rm tot}\right)^2$ is the total variance in $D_H$ given as

\begin{equation}
\left( \Delta D_H^{\rm tot}(z_{\rm BAO},M_B,W_{\rm k0},r_d) \right)^2 = \left(\Delta D_H^{\rm BAO1}\right)^2 +\left( \Delta D_H^{\rm GPR}(z_{\rm BAO},M_B,W_{\rm k0},r_d) \right)^2 .
\end{equation}

Next, using the reconstructed $m_{\rm GPR}$ (obtained from Sec.~\ref{sec-SN_GPR}), we compute $D_M$ (via $d_L$) at BAO redshift points using Eq.~\eqref{eq:DM_from_dL} and the corresponding uncertainties. We denote these as the $D_{M}^{\rm GPR}$ and $\Delta D_{M}^{\rm GPR}$ respectively. We denote the $D_M$ and $\Delta D_M$ directly obtained from the BAO2 data as $D_{M}^{\rm BAO2}$ and $\Delta D_{M}^{\rm BAO2}$ respectively. We compare these two by defining a corresponding log-likelihood ($\log{L}_{\rm BAO2}$) given as

\begin{eqnarray}
\log{L}_{\rm BAO2} (M_B,r_d) &=& - \frac{1}{2} \sum_{z_{\rm BAO}} \frac{ \left[ D_M^{\rm GPR}(z_{\rm BAO},M_B,r_d)-D_M^{\rm BAO2} \right]^2}{ \left(\Delta D_M^{\rm tot}(z_{\rm BAO},M_B,r_d)\right)^2 } \nonumber\\
&& - \frac{1}{2} \sum_{z_{\rm BAO}} \log{ \left[ 2 \pi \left(\Delta D_M^{\rm tot}(z_{\rm BAO},M_B,r_d)\right)^2 \right] },
\label{eq:lnlk_SN_BAO_2}
\end{eqnarray}

\noindent
where $(\Delta D_M^{\rm tot})^2$ is the total variance in $D_M$ given as

\begin{equation}
\left(\Delta D_M^{\rm tot}(z_{\rm BAO},M_B,r_d)\right)^2 = \left(\Delta D_M^{\rm BAO2}\right)^2 + \left(\Delta D_M^{\rm GPR}(z_{\rm BAO},M_B,r_d)\right)^2 .
\end{equation}

\noindent
Note that, in this step, the $W_{\rm k0}$ parameter is not involved. That is why, in the above two equations, the $W_{\rm k0}$ parameter is not present.

So, the total log-likelihood for the two kinds of BAO observations, $\log{L}_{\rm BAO}$ is given as

\begin{equation}
\log{L}_{\rm BAO} (M_B,W_{\rm k0},r_d) = \log{L}_{\rm BAO1} (M_B,W_{\rm k0},r_d) + \log{L}_{\rm BAO2} (M_B,r_d) .
\label{eq:lnlk_SN_BAO}
\end{equation}

Here, one important fact to notice is that the $r_d$ parameter is degenerate to both $M_B$ and $W_{\rm k0}$. This can be seen through Eqs.~\eqref{eq:Hubble},~\eqref{eq:dL_from_m},~\eqref{eq:dLp_from_mp},~\eqref{eq:DH_from_H}, and~\eqref{eq:DM_from_dL}. So, to break these degeneracies, we have to combine the CC data to the SN and BAO data. So, we get simultaneous constraints on $M_B$, $W_{\rm k0}$, and $r_d$ from combinations of SN, CC, and BAO data by maximizing the corresponding log-likelihood ($\log{L}_{\rm CC+BAO}$) given as

\begin{equation}
\log{L}_{\rm CC+BAO} (M_B,W_{\rm k0},r_d) = \log{L}_{\rm CC} (M_B,W_{\rm k0}) +\log{L}_{\rm BAO} (M_B,W_{\rm k0},r_d).
\label{eq:lnlk_CC_SN_BAO}
\end{equation}

\noindent
Consequently we get simultaneous constraints on $M_B$, $\Omega_{\rm k0}h^2$, and $r_d$. Note that, since $r_d$ is degenerate to both $M_B$ and $\Omega_{\rm k0}h^2$ (through $W_{\rm k0}$), addition of BAO data to SN+CC data makes the constraints on $M_B$ and $\Omega_{\rm k0}h^2$ tighter while simultaneously gives constraints on $r_d$.

\subsection{Combination of SN, CC, BAO, and GRB data: bounds on gamma-ray bursts nuisance parameters}
\label{sec-SN_GRB}

The Amati correlated gamma-ray bursts (GRB) data have simultaneous measurements in the observed peak energy, $E_p^{\rm obs}$ of GRB photons and bolometric fluence, $S_{\rm bolo}$ \cite{Khadka:2021vqa,Wei:2008kq,Amati:2008hq}. These quantities are related to the luminosity distance through Amati relations given by

\begin{eqnarray}
&& E_{\rm iso} = 4 \pi d_L^2 S_{\rm bolo} (1+z)^{-1}, 
\label{eq:Amati_relation_2} \\
&& P = \log_{\rm 10}{\left(\frac{E_{\rm iso}}{{\rm erg}}\right)},
\label{eq:Amati_relation_1_defn} \\
&& E_p^{\rm obs} = \frac{E_p}{1+z},
\label{eq:Amati_relation_3} \\
&& P = a+b \log_{\rm 10}{\left(\frac{E_p^{\rm obs}}{{\rm keV}}\right)},
\label{eq:Amati_relation_1}
\end{eqnarray}

\noindent
where $E_{\rm iso}$ is the isotropic energy and $E_p$ is the rest-frame peak energy of GRB photons; $a$ and $b$ are two nuisance parameters involved in the GRB observations.

From the reconstructed $m_{\rm GRB}$ and $\Delta m_{\rm GRB}$ (obtained from Sec.~\ref{sec-SN_GPR}), we compute $d_L$ and $\Delta d_L$ at GRB redshift points using Eq.~\eqref{eq:dL_from_m}. Note that these values are functions of $M_B$ parameter which can be seen from Eq.~\eqref{eq:dL_from_m}. Next, from the reconstructed $d_L$ and $\Delta d_L$, we compute $E_{\rm iso}$ and $\Delta E_{\rm iso}$ using Eq.~\eqref{eq:Amati_relation_2}. Note that, in this step, we use the $S_{\rm bolo}$ data from GRB observations. Next, from these $E_{\rm iso}$ and $\Delta E_{\rm iso}$, we compute $P$ and $\Delta P$ using Eq.~\eqref{eq:Amati_relation_1_defn}. We denote these as $P_{\rm GPR}$ and $\Delta P_{\rm GPR}$. Note that these are dependent on only $M_B$ parameter.

From $E_p^{\rm obs}$ data from GRB observations, we compute $E_p$ and $\Delta E_p$ using Eq.~\eqref{eq:Amati_relation_3}. From these $E_p$ and $\Delta E_p$, we compute $P$ and $\Delta P$ using Eq.~\eqref{eq:Amati_relation_1}. Note that these are functions of $a$ and $b$ parameters which can be seen through Eq.~\eqref{eq:Amati_relation_1}. We denote this as $P_{\rm GRB}$ and $\Delta P_{\rm GRB}$ respectively. Note that, in Eq.~\eqref{eq:Amati_relation_1}, $P$ is linear in $a$ and $b$, and only $b$ is multiplied by an uncertainty propagating term but not in the case of $a$. So, in the uncertainty propagation through Eq.~\eqref{eq:Amati_relation_1}, the parameter $a$ is not propagated in $\Delta P_{\rm GRB}$. So, $\Delta P_{\rm GRB}$ is the function of only $b$ parameter but $P_{\rm GRB}$ is the function of both $a$ and $b$ parameters.

The total variance in $P$ is enhanced by an extra term given as $\sigma_{\rm ext}^2$, where $\sigma_{\rm ext}$ is a dispersion parameter. We now compare $P_{\rm GPR}$ with $P_{\rm GRB}$ by defining a log-likelihood ($\log{L}_{\rm GRB}$) defined as

\begin{eqnarray}
\log{L}_{\rm GRB} (M_B,a,b,\sigma_{\rm ext}) &=& - \frac{1}{2} \sum_{z_{\rm GRB}} \frac{ \left[ P_{\rm GPR}(z_{\rm GRB},M_B)-P_{\rm GRB}(a,b) \right]^2}{ \Delta P_{\rm tot}^2(z_{\rm GRB},M_B,b) +\sigma_{\rm ext}^2 } \nonumber\\
&& - \frac{1}{2} \sum_{z_{\rm GRB}} \log{ \left[ 2 \pi \left(\Delta P_{\rm tot}^2(z_{\rm GRB},M_B,b) +\sigma_{\rm ext}^2\right) \right] },
\label{eq:lnlk_SN_GRB}
\end{eqnarray}

\noindent
where $\Delta P_{\rm tot}^2$ is given as $\Delta P_{\rm tot}^2(z_{\rm GRB},M_B,b)=\Delta P_{\rm GPR}^2(z_{\rm GRB},M_B)+\Delta P_{\rm GRB}^2(b)$.

One can check that, here, $M_B$ and $a$ parameters are degenerate. To break this degeneracy, we have to add the CC data, and the corresponding log-likelihood ($\log{L}_{\rm CC+GRB}$) is given as

\begin{equation}
\log{L}_{\rm CC+GRB} (M_B,W_{\rm k0},a,b,\sigma_{\rm ext}) = \log{L}_{\rm CC} (M_B,W_{\rm k0}) +\log{L}_{\rm GRB} (M_B,a,b,\sigma_{\rm ext}).
\label{eq:lnlk_SN_CC_GRB}
\end{equation}

We also combine BAO data for consistency checks. For this case, we get simultaneous constraints on $M_B$, $W_{\rm k0}$, $r_d$, $a$, $b$, and $\sigma_{\rm ext}$ by maximizing the corresponding log-likelihood ($\log{L}_{\rm CC+BAO+GRB}$) given as

\begin{eqnarray}
\log{L}_{\rm CC+BAO+GRB} (M_B,W_{\rm k0},r_d,a,b,\sigma_{\rm ext}) &=& \log{L}_{\rm CC+BAO} (M_B,W_{\rm k0},r_d) \nonumber\\
&& +\log{L}_{\rm GRB} (M_B,a,b,\sigma_{\rm ext}).
\label{eq:lnlk_CC_SN_BAO_GRB}
\end{eqnarray}

\begin{figure}
\centering
\includegraphics[width=1.0\textwidth]{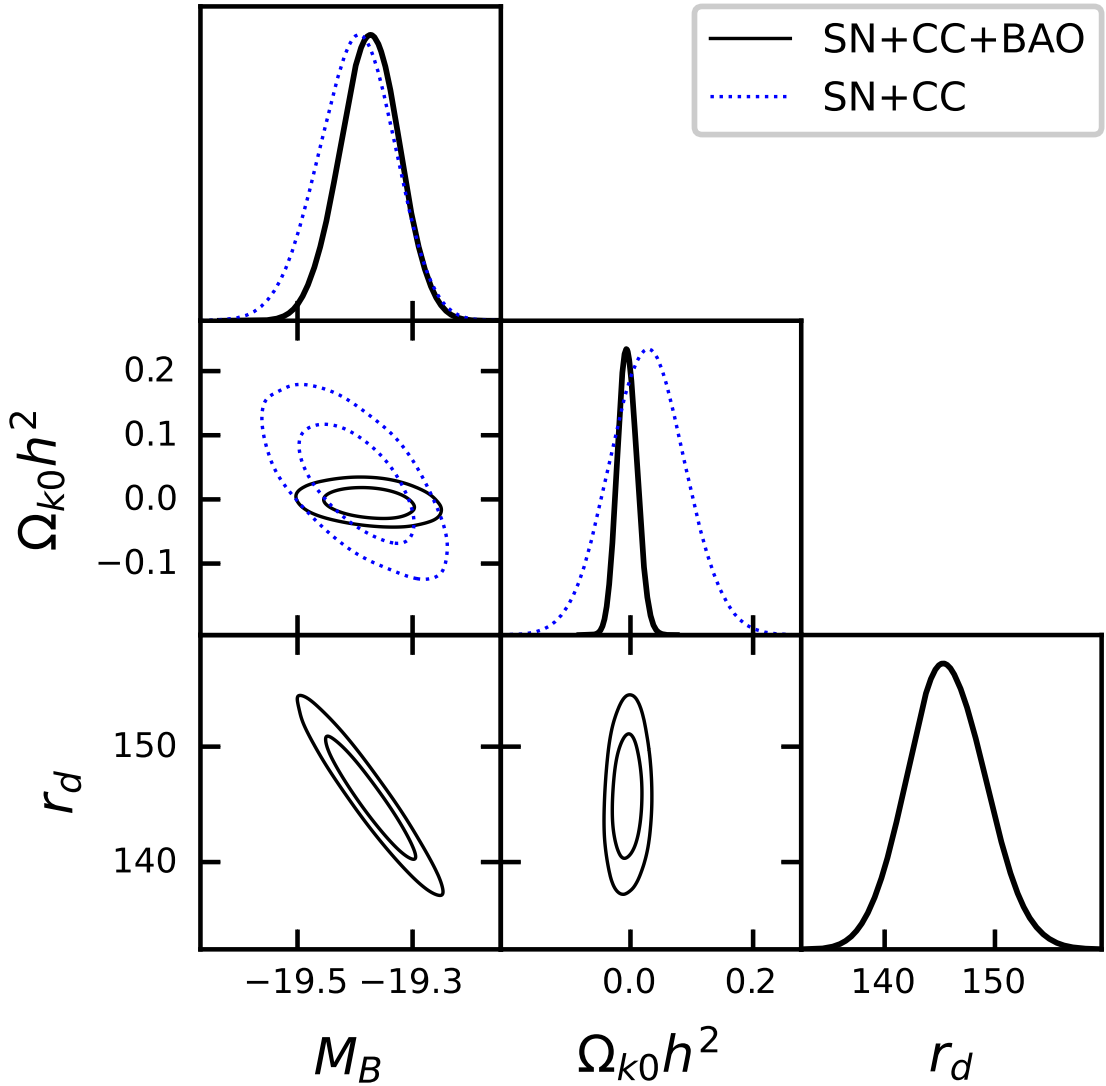}
\caption{
\label{fig:bounds_main}
Constraints on $M_B$, $\Omega_{\rm k0}h^2$, and $r_d$ parameters obtained from two different combinations of data. For a particular color or type, the inner and the outer contours correspond to the 1$\sigma$ and 2$\sigma$ confidence contours respectively. Dotted-blue and solid-black contours are obtained from SN+CC and SN+CC+BAO combinations of data respectively.
}
\end{figure}

\begin{figure*}
\centering
\includegraphics[width=1.0\textwidth]{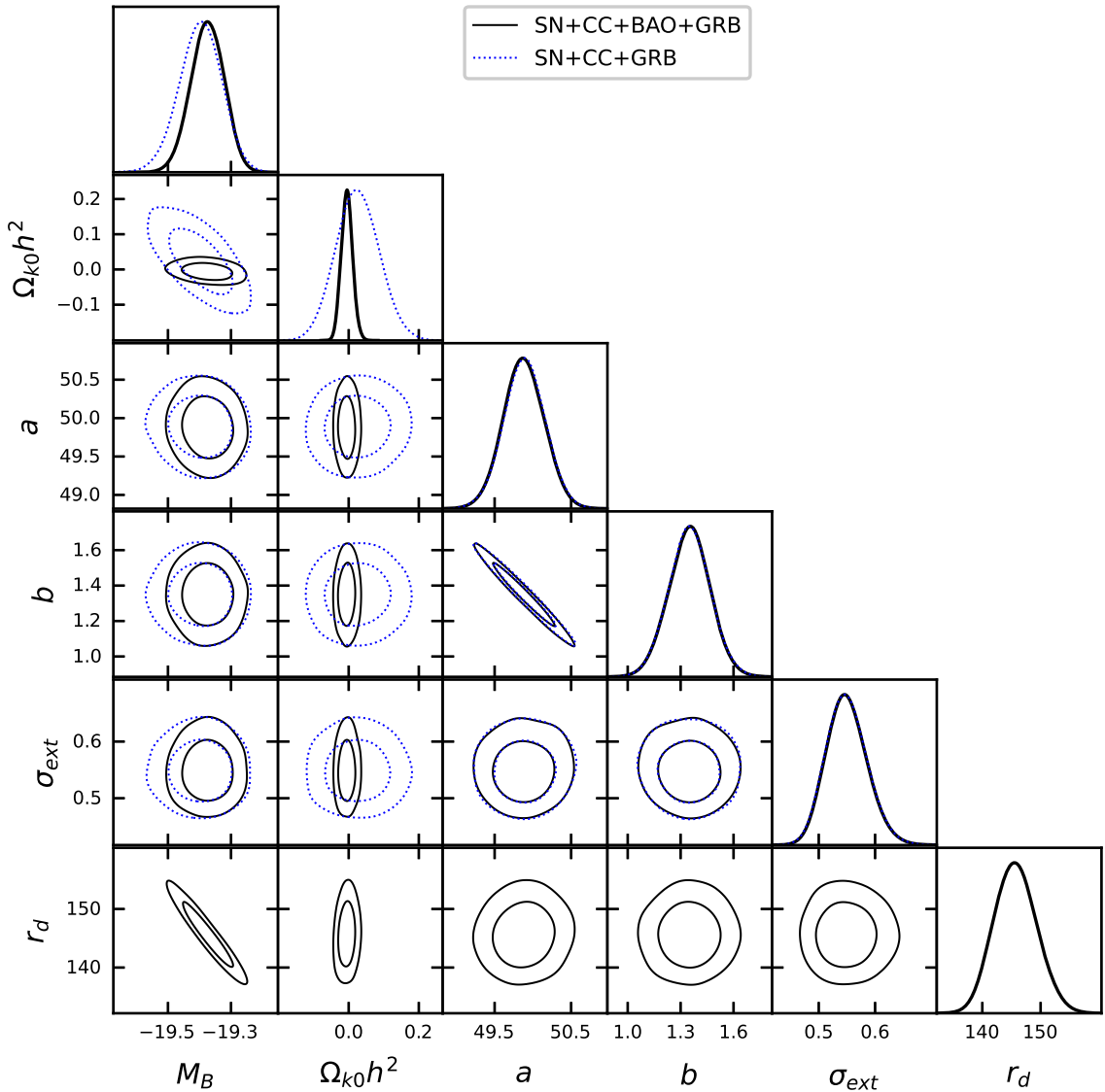}
\caption{
\label{fig:bounds_GRBs}
Constraints on $M_B$, $\Omega_{\rm k0}h^2$, $r_d$, and GRB data-related nuisance parameters, $a$, $b$, and $\sigma_{\rm ext}$ obtained from SN+CC+GRB and SN+CC+BAO+GRB combinations of data. For a particular color or type, the inner and the outer contours correspond to the 1$\sigma$ and 2$\sigma$ confidence contours respectively. Dotted-blue and solid-black contours are obtained from SN+CC+GRB and SN+CC+BAO+GRB combinations of data respectively.
}
\end{figure*}

\begin{figure}
\centering
\includegraphics[width=1.0\textwidth]{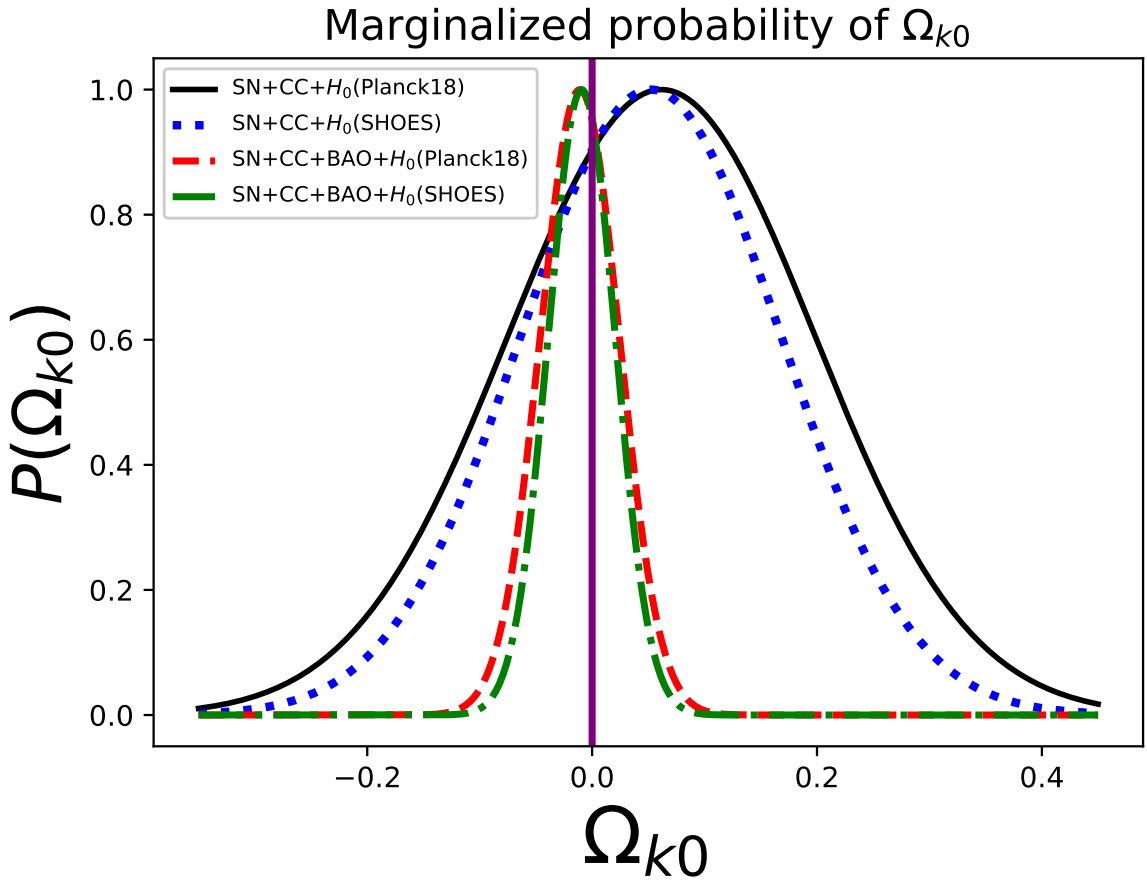}
\caption{
\label{fig:bounds_Ok0}
Marginalized probability distribution of $\Omega_{\rm k0}$ corresponding to different combinations of data.
}
\end{figure}

\section{Results}
\label{sec-results}

The constraints on the nuisance parameters are dependent on the constraints on the $W_{\rm k0}$ parameter that is related to $\Omega_{\rm k0}h^2$ through Eq.~\eqref{eq:defnWk0}. We quote all the results through $\Omega_{\rm k0}h^2$ parameter instead of $W_{\rm k0}$.

\begin{table}
\begin{center}
\begin{tabular}{ |c|c|c|  }
\hline
Parameters & SN+CC & SN+CC+BAO \\
\hline
$M_B$ & $-19.398\pm 0.066$ & $-19.375\pm 0.052$ \\
\hline
$\Omega_{\rm k0}h^2$ & $0.028\pm0.062$ & $-0.005\pm0.016$ \\
\hline
$r_d$ & - & $145.6\pm 3.5$ \\
\hline
\end{tabular}
\end{center}
\caption{
Constraints on $M_B$, $\Omega_{\rm k0}h^2$, and $r_d$ parameters obtained from SN+CC and SN+CC+BAO combinations of data.
}
\label{table:main}
\end{table}

In Figure~\ref{fig:bounds_main}, we have shown constraints on $M_B$, $\Omega_{\rm k0}h^2$, and $r_d$ parameters obtained from two different combinations of data. The dotted-blue and solid-black lines correspond to the constraints obtained from SN+CC and SN+CC+BAO combinations of data respectively. For SN+CC, we obtain constraints only on $M_B$ and $\Omega_{\rm k0}h^2$ not on $r_d$. The addition of BAO data put constraints on $r_d$ along with $M_B$ and $\Omega_{\rm k0}h^2$. For this case, constraints on $M_B$ and $\Omega_{\rm k0}h^2$ are comparatively tighter. The values of these constraints are mentioned in Table~\ref{table:main}.

\begin{table}
\begin{center}
\begin{tabular}{ |c|c|c|  }
\hline
Parameters & SN+CC+GRB & SN+CC+BAO+GRB \\
\hline
$M_B$ & $-19.397\pm 0.067$ & $-19.375\pm 0.053$ \\
\hline
$\Omega_{\rm k0}h^2$ & $0.025\pm0.062$ & $-0.005\pm0.016$ \\
\hline
$r_d$ & - & $145.7^{+3.4}_{-3.8}$ \\
\hline
$a$ & $49.89\pm 0.27$ & $49.88\pm 0.27$ \\
\hline
$b$ & $1.35\pm 0.12$ & $1.35\pm 0.12$ \\
\hline
$\sigma_{\rm ext}$ & $0.550^{+0.033}_{-0.039}$ & $0.550^{+0.033}_{-0.038}$ \\
\hline
\end{tabular}
\end{center}
\caption{
Constraints on $M_B$, $\Omega_{\rm k0}h^2$, $r_d$, $a$, $b$, and $\sigma_{\rm ext}$ parameters obtained from SN+CC+GRB and SN+CC+BAO+GRB combinations of data.
}
\label{table:GRB}
\end{table}

In Figure~\ref{fig:bounds_GRBs}, we have shown the constraints on the GRB nuisance parameters, $a$,  $b$, and $\sigma_{\rm ext}$ along with $M_B$, $\Omega_{\rm k0}h^2$, and $r_d$ parameters obtained from SN+CC+GRB and SN+CC+BAO+GRB combinations of data. For a particular color or type, the inner and the outer contours correspond to the 1$\sigma$ and 2$\sigma$ confidence contours respectively. Dotted-blue and solid-black contours are obtained from SN+CC+GRB and SN+CC+BAO+GRB combinations of data respectively. Note that, the constraints on the $M_B$, $\Omega_{\rm k0}h^2$, and $r_d$ are similar as in Figure~\ref{fig:bounds_main}. This can be seen in Table~\ref{table:GRB} by the comparison in Table~\ref{table:main}. This indicates that the constraints on $M_B$, $\Omega_{\rm k0}h^2$, and $r_d$ parameters are almost unaffected by the addition of the GRB data.

From all the figures and tables, we can see that the constraints on the GRB nuisance parameters are almost independent of the constraints on the $M_B$, $\Omega_{\rm k0}h^2$, and $r_d$.

Another important fact to notice is that constraints on all the nuisance parameters are dependent on $\Omega_{\rm k0} h^2$ (or equivalently on $\Omega_{\rm k0} H_0^2$) but not individually on each of $\Omega_{\rm k0}$ and $H_0$. To get constraints on $\Omega_{\rm k0}$ and $H_0$ parameters individually, one has to add data that directly provide observations either in $H_0$ or in the $\Omega_{\rm k0}$ separately. That is why next we consider $H_0$ priors to get constraints on $\Omega_{\rm k0}$. For this purpose, we consider two kinds of priors on $H_0$. One is from the results of the Planck 2018 mission \citep{Planck:2018vyg} and another one is from the SHOES \citep{Riess:2020fzl} observations. These are mentioned in Table~\ref{table:H0_list}.

\begin{table}
\begin{center}
\begin{tabular}{ |c|c|c|  }
\hline
Planck18 & $H_0 = 67.4\pm0.5$ & \citep{Planck:2018vyg} \\
\hline
SHOES & $H_0 = 73.2\pm1.3$ & \citep{Riess:2020fzl} \\
\hline
\end{tabular}
\end{center}
\caption{
$H_0$ priors from Planck 18 mission and SHOES observations. The values are mentioned in KmS$^{-1}$Mpc$^{-1}$ unit.
}
\label{table:H0_list}
\end{table}

From $\Omega_{\rm k0}h^2$ and $H_0$, we compute $\Omega_{\rm k0}$ by using the equation given as

\begin{equation}
\Omega_{\rm k0} = \left( \Omega_{\rm k0}h^2 \right) \left[ \frac{ 100 \hspace{0.1cm} \text{Km S}^{-1} \text{Mpc}^{-1} }{H_0} \right]^2 .
\end{equation}

\noindent
And we compute the corresponding uncertainties using the propagataion of uncertainty. We have listed the constraints on $\Omega_{\rm k0}$ for different cases in Table~\ref{table:Ok0} and plotted the 1D marginalized probability distribution of $\Omega_{\rm k0}$ in Figure~\ref{fig:bounds_Ok0}, obtained from different combinations of data, mentioned in the figure. We have not shown the constraints from the GRB data, because the addition of these data does not affect the constraints on the $\Omega_{\rm k0}h^2$ parameter and consequently on $\Omega_{\rm k0}$. We have already seen this fact from previous figures and tables. This is obvious because $\log{L}_{\rm GRB}$ in Eq.~\eqref{eq:lnlk_SN_GRB} is independent of $W_{\rm k0}$ parameter.

\begin{table}
\begin{center}
\begin{tabular}{ |c|c|c|  }
\hline
Data combinations & +$H_0$(Planck18) & +$H_0$(SHOES) \\
\hline
SN+CC & $0.062\pm0.136$ & $0.052\pm0.116$ \\
\hline
SN+CC+BAO & $-0.011 \pm 0.035$ & $-0.009 \pm 0.030$ \\
\hline
\end{tabular}
\end{center}
\caption{
Constraints on $\Omega_{\rm k0}$ parameter obtained from different combinations of data.
}
\label{table:Ok0}
\end{table}

We see that the mean values of $\Omega_{\rm k0}$ are well inside the 1$\sigma$ confidence region. So, there are no significant deviations from a flat Universe.

\section{Conclusion}
\label{sec-conclusion}

This analysis concludes that we can get constraints on cosmological nuisance parameters and the cosmic curvature density parameter corresponding to different observations by the different combinations of data in a minimal model dependent way using Gaussian process regression (GPR) analysis. This analysis has been done for important observations like SN, CC, BAO, GRB, and $H_0$ observations from Planck 2018 mission and SHOES. The constraints on the nuisance parameters, obtained in this analysis, can be used as the prior for the cosmological data analysis.

The results, obtained in this analysis, are not completely model independent, but model dependence is minimal. This is because there is model dependence through the kernel and the mean function in the GPR analysis and through data like BAO which consider a particular fiducial model. Hence the methodology, presented here, has minimal model dependence. The model dependence is not very significant though. In the future, the GPR analysis can be replaced by more accurate, advanced, and modern reconstruction methods like deep learning and neural networks, but the methodology, presented here, would be very helpful to do that.

\appendix
\section{Gaussian process regression analysis}
\label{sec-gpr_basic}

Let us consider we have $n$ number of data points from a kind of observation. $X$, $Y$, and $\Delta Y$ are vectors of observational data points, the corresponding mean values of a quantity, and the standard deviation values of that quantity respectively. Gaussian process regression (GPR) analysis can predict the mean and standard deviation values of the same quantity at some different target points. Let us denote $X_*$, $Y_*$, and $\Delta Y_*$ are vectors of target points, the corresponding mean values, and standard deviation values respectively with the help of a kernel covariance function and a mean function. Using GPR, we get the predicted values given as \citep{williams1995gaussian,GpRasWil,Seikel_2012,Shafieloo_2012}

\begin{align}
\label{eq:GPR_mean_prediction}
Y_* &= M_* + K(X_*,X) \left[ K(X,X)+C \right]^{-1} (Y-M), \\
\label{eq:GPR_std_prediction}
\text{Cov}[Y_*,Y_*] &= K(X_*,X_*) - K(X_*,X) \left[ K(X,X)+C \right]^{-1} K(X,X_*),
\end{align}

\noindent
where 'Cov' stands for covariance, $K$ is the kernel matrix according to a particular kernel covariance function, $M$ and $M_*$ are the mean vectors at data points and target points respectively corresponding to a particular mean function, and $C$ is the noise covariance matrix of the data. If errors in data are uncorrelated then an element, $C_{ij}$ of $C$ matrix would be $C_{ij} = (\Delta Y_i)^2 \delta_{ij}$, with $\delta_{ij}$ being the usual Kronecker delta. If the errors in data are correlated then off-diagonal elements of $C$ would be non-zero accordingly.

The details of the kernel covariance and mean functions are briefly discussed in the next section.

The kernel and mean functions have some parameters. We have to take their best-fit values for the mean prediction in Eqs.~\eqref{eq:GPR_mean_prediction} and~\eqref{eq:GPR_std_prediction}. To do so, we minimize the negative of log marginal likelihood (denoted by $\log P(Y|X)$) given as \citep{Seikel_2012}

\begin{eqnarray}
\log P(Y|X) &=& -\frac{1}{2} (Y-M)^T \left[ K(X,X)+C \right]^{-1} (Y-M) \nonumber\\
&& -\frac{1}{2} \log |K(X,X)+C| -\frac{n}{2} \log{(2 \pi)},
\label{eq:log_marginal_likelihood}
\end{eqnarray}

\noindent
where $|K(X,X)+C|$ is the determinant of the $K(X,X)+C$ matrix.

GPR can also predict the gradient of a quantity, for example, here, $y'=\frac{dy}{dx}$. By prime notation, we mean the derivative of a quantity w.r.t the argument, for example, here $x$, and in the main text, it is $z$. From GPR predictions, the mean vector and the covariance matrix corresponding to the first derivative are given as \citep{Seikel_2012}

\begin{align}
\label{eq:derivative_mean_predictions}
Y'_* &= M'_* + [K'(X,X_*)]^T \left[ K(X,X)+C \right]^{-1} (Y-M), \\
\label{eq:derivative_std_predictions}
\text{Cov}[Y'_*,Y'_*] &= K''(X_*,X_*) - [K'(X,X_*)]^T \left[ K(X,X)+C \right]^{-1} K'(X,X_*),
\end{align}

\noindent
where prime and double prime are the first and second derivatives of the corresponding function respectively; $k'(x,x_*)$ and $k''(x_*,x_*)$ are given as

\begin{eqnarray}
k'(x,x_*) = \dfrac{\partial k(x,x_*)}{\partial x_*}, \hspace{1 cm}
k''(x_*,x_*) = \dfrac{\partial ^2 k(x_*,x_*)}{\partial x_* \partial x_*},
\label{eq:kernel_derivatives}
\end{eqnarray}

\noindent
respectively. Here, by the notation $k$, we denote the matrix element of the main matrix $K$.

\begin{table*}
\begin{center}
\begin{tabular}{ |c|c|c|c|c|c|  }
\hline
Parameters & SE: $\Lambda$CDM & M5by2: $\Lambda$CDM & RQ: $\Lambda$CDM & SE: wCDM & SE: CPL \\
\hline
$M_B$ & $-19.375\pm 0.053$ & $-19.374\pm 0.051$ & $-19.374\pm 0.051$ & $-19.374\pm 0.052$ & $-19.373\pm 0.052$ \\
\hline
$\Omega_{\rm k0}h^2$ & $-0.005\pm0.016$ & $-0.008\pm0.016$ & $-0.006\pm0.016$ & $-0.007\pm0.016$ & $-0.006\pm0.017$ \\
\hline
$r_d$ & $145.7^{+3.4}_{-3.8}$ & $145.5\pm3.5$ & $145.5\pm3.5$ & $145.6\pm3.5$ & $145.5\pm3.5$ \\
\hline
$a$ & $49.88\pm 0.27$ & $49.88\pm 0.27$ & $49.89\pm 0.27$ & $49.89\pm 0.27$ & $49.88\pm 0.27$ \\
\hline
$b$ & $1.35\pm 0.12$ & $1.35\pm 0.12$ & $1.35\pm 0.12$ & $1.34\pm 0.12$ & $1.35\pm 0.12$ \\
\hline
$\sigma_{\rm ext}$ & $0.550^{+0.033}_{-0.038}$ & $0.548^{+0.032}_{-0.039}$ & $0.550^{+0.032}_{-0.039}$ & $0.551^{+0.032}_{-0.040}$ & $0.548^{+0.031}_{-0.039}$ \\
\hline
\end{tabular}
\end{center}
\caption{
Constraints on $M_B$, $\Omega_{\rm k0}h^2$, $r_d$, $a$, $b$, and $\sigma_{\rm ext}$ parameters obtained from SN+CC+BAO+GRB data for different kernels and means.
}
\label{table:GRB_dfrnt}
\end{table*}

\section{Dependence of GPR predictions on kernels and mean functions}
\label{sec-parameters_dependence}

Some popular kernel covariance functions are listed below:

\begin{align}
\label{eq:kernel_SE}
k(d) &= \sigma_f^2 e^{ -\frac{d^2}{2 l^2} } \hspace{0.2 cm} \text{(squared exponential)}, \\
\label{eq:Matern}
k(d) &= \sigma_f^2  \left( 1+\frac{\sqrt{5}d}{l}+\frac{5d^2}{3l^2} \right) e^{ -\frac{\sqrt{5}d}{l} } \hspace{0.2 cm} \text{(Mat\'ern with order 5/2)}, \\
\label{eq:rational_quadratic_RQ}
k(d) &= \sigma_f^2  \left( 1+\frac{d^2}{2 r l^2} \right)^{-r} \hspace{0.2 cm} \text{(rational quadratic)},
\end{align}

\noindent
where $\sigma_f$, $l$, $r$ are the corresponding kernel parameters which are called the hyperparameters; $d=|x_1-x_2|$ and $k(x_1,x_2)=k(|x_1-x_2|)=k(d)$. Among the above-listed kernels, the first one i.e. the squared exponential kernel is used in the main text. We denote this kernel as 'SE'. The second kernel is the Mat\'ern kernel covariance function with order $5/2$. We denote this kernel as 'M5by2'. The third one is the rational quadratic kernel covariance function. We denote this as 'RQ'.

In the main text, we have considered the $\Lambda$CDM model for the mean function for $m(z)$. Here we also consider the more general model, called Chevallier-Polarski-Linder (CPL) parametrization \citep{Chevallier:2000qy,Linder:2002et}, where the equation of state of the dark energy is given as

\begin{equation}
w = w_0+w_a \frac{z}{1+z},
\label{eq:w_DE_CPL}
\end{equation}

\noindent
where $w_0$ and $w_a$ are two model parameters. In the CPL model, the Hubble parameter is given as

\begin{equation}
\frac{H^2 }{H_0^2} = \Omega_{\rm m0}(1+z)^{3}+\Omega_{\rm k0}(1+z)^{2} +(1-\Omega_{\rm m0}-\Omega_{\rm k0}) (1+z)^{3(1+w_{0}+w_{a})} e^{-\frac{3w_{a}z}{(1+z)}},
\label{eq:DE_Hsqr}
\end{equation}

\noindent
where $\Omega_{\rm m0}$ is the matter-energy density parameter. wCDM model is the subset of CPL model where $w_a=0$ and $\Lambda$CDM model is the further subset where $w_a=0$ and $w_0=-1$.

Using Eq.~\eqref{eq:DE_Hsqr}, we can compute any quantity from Eq.~\eqref{eq:loscov} to Eq.~\eqref{eq:dL_to_m_basic} for $\Lambda$CDM, wCDM and CPL models accordingly.

To show how our results depend on different kernel and mean functions, in Table~\ref{table:GRB_dfrnt}, we list constraints on the parameters for SN+CC+BAO+GRB combinations of data for different kernel and mean functions. We have not included other combinations of data, because only this combination of data is enough to show the fact that the dependence of the constraints on different kernels and mean functions is not significant.

\acknowledgments
The author would like to acknowledge IISER Kolkata for the financial support through the postdoctoral fellowship.


\bibliographystyle{plain}
\bibliography{refs}

\begin{thebibliography}{10}

\bibitem{Planck:2013pxb}
P.~A.~R. Ade et~al.
\newblock {Planck 2013 results. XVI. Cosmological parameters}.
\newblock {\em Astron. Astrophys.}, 571:A16, 2014.

\bibitem{Planck:2015fie}
P.~A.~R. Ade et~al.
\newblock {Planck 2015 results. XIII. Cosmological parameters}.
\newblock {\em Astron. Astrophys.}, 594:A13, 2016.

\bibitem{Planck:2018vyg}
N.~Aghanim et~al.
\newblock {Planck 2018 results. VI. Cosmological parameters}.
\newblock {\em Astron. Astrophys.}, 641:A6, 2020.
\newblock [Erratum: Astron.Astrophys. 652, C4 (2021)].

\bibitem{BOSS:2016wmc}
Shadab Alam et~al.
\newblock {The clustering of galaxies in the completed SDSS-III Baryon
  Oscillation Spectroscopic Survey: cosmological analysis of the DR12 galaxy
  sample}.
\newblock {\em Mon. Not. Roy. Astron. Soc.}, 470(3):2617--2652, 2017.

\bibitem{eBOSS:2020yzd}
Shadab Alam et~al.
\newblock {Completed SDSS-IV extended Baryon Oscillation Spectroscopic Survey:
  Cosmological implications from two decades of spectroscopic surveys at the
  Apache Point Observatory}.
\newblock {\em Phys. Rev. D}, 103(8):083533, 2021.

\bibitem{Amati:2008hq}
Lorenzo Amati, Cristiano Guidorzi, Filippo Frontera, Massimo Della~Valle, Fabio
  Finelli, Raffaella Landi, and Enrico Montanari.
\newblock {Measuring the cosmological parameters with the Ep,i-Eiso correlation
  of Gamma-Ray Bursts}.
\newblock {\em Mon. Not. Roy. Astron. Soc.}, 391:577--584, 2008.

\bibitem{Barboza:2008rh}
E.~M. Barboza, Jr. and J.~S. Alcaniz.
\newblock {A parametric model for dark energy}.
\newblock {\em Phys. Lett. B}, 666:415--419, 2008.

\bibitem{Bull:2015stt}
Philip Bull et~al.
\newblock {Beyond $\Lambda$CDM: Problems, solutions, and the road ahead}.
\newblock {\em Phys. Dark Univ.}, 12:56--99, 2016.

\bibitem{Cai:2015pia}
Rong-Gen Cai, Zong-Kuan Guo, and Tao Yang.
\newblock {Null test of the cosmic curvature using $H(z)$ and supernovae data}.
\newblock {\em Phys. Rev. D}, 93(4):043517, 2016.

\bibitem{Camarena:2019rmj}
David Camarena and Valerio Marra.
\newblock {A new method to build the (inverse) distance ladder}.
\newblock {\em Mon. Not. Roy. Astron. Soc.}, 495(3):2630--2644, 2020.

\bibitem{Camarena:2021jlr}
David Camarena and Valerio Marra.
\newblock {On the use of the local prior on the absolute magnitude of Type Ia
  supernovae in cosmological inference}.
\newblock {\em Mon. Not. Roy. Astron. Soc.}, 504:5164--5171, 2021.

\bibitem{Cao:2022ugh}
Shulei Cao and Bharat Ratra.
\newblock {Using lower-redshift, non-CMB, data to constrain the Hubble constant
  and other cosmological parameters}.
\newblock 3 2022.

\bibitem{Carroll:2000fy}
Sean~M. Carroll.
\newblock {The Cosmological constant}.
\newblock {\em Living Rev. Rel.}, 4:1, 2001.

\bibitem{Camlibel:2020xbn}
A.~K. \c{C}aml\i{}bel, \.I. Semiz, and M.~A. Feyizo\u{g}lu.
\newblock {Pantheon update on a model-independent analysis of cosmological
  supernova data}.
\newblock {\em Class. Quant. Grav.}, 37(23):235001, 2020.

\bibitem{Chevallier:2000qy}
Michel Chevallier and David Polarski.
\newblock {Accelerating universes with scaling dark matter}.
\newblock {\em Int. J. Mod. Phys. D}, 10:213--224, 2001.

\bibitem{Clifton:2011jh}
Timothy Clifton, Pedro~G. Ferreira, Antonio Padilla, and Constantinos Skordis.
\newblock {Modified Gravity and Cosmology}.
\newblock {\em Phys. Rept.}, 513:1--189, 2012.

\bibitem{Colgain:2022nlb}
Eoin~\'O. Colg\'ain, M.~M. Sheikh-Jabbari, Rance Solomon, Giada Bargiacchi,
  Salvatore Capozziello, Maria~Giovanna Dainotti, and Dejan Stojkovic.
\newblock {Revealing Intrinsic Flat $\Lambda$CDM Biases with Standardizable
  Candles}.
\newblock 3 2022.

\bibitem{Copeland:2006wr}
Edmund~J. Copeland, M.~Sami, and Shinji Tsujikawa.
\newblock {Dynamics of dark energy}.
\newblock {\em Int. J. Mod. Phys. D}, 15:1753--1936, 2006.

\bibitem{DiValentino:2021izs}
Eleonora Di~Valentino, Olga Mena, Supriya Pan, Luca Visinelli, Weiqiang Yang,
  Alessandro Melchiorri, David~F. Mota, Adam~G. Riess, and Joseph Silk.
\newblock {In the Realm of the Hubble tension $-$ a Review of Solutions}.
\newblock 3 2021.

\bibitem{Dinda:2019mev}
Bikash~R. Dinda.
\newblock {Model independent parametrization of the late time cosmic
  acceleration: Constraints on the parameters from recent observations}.
\newblock {\em Phys. Rev. D}, 100(4):043528, 2019.

\bibitem{Dinda:2021ffa}
Bikash~R. Dinda.
\newblock {Cosmic expansion parametrization: Implication for curvature and H0
  tension}.
\newblock {\em Phys. Rev. D}, 105(6):063524, 2022.

\bibitem{Dinda:2022jih}
Bikash~R. Dinda and Narayan Banerjee.
\newblock {Model independent bounds on Type Ia supernova absolute peak
  magnitude}.
\newblock 8 2022.

\bibitem{Escamilla-Rivera:2019hqt}
Celia Escamilla-Rivera, Maryi Alejandra~Carvajal Quintero, and S.~Capozziello.
\newblock {A deep learning approach to cosmological dark energy models}.
\newblock {\em JCAP}, 03:008, 2020.

\bibitem{Gomez-Valent:2021hda}
Adri\`a G\'omez-Valent.
\newblock {Measuring the sound horizon and absolute magnitude of SNIa by
  maximizing the consistency between low-redshift data sets}.
\newblock {\em Phys. Rev. D}, 105(4):043528, 2022.

\bibitem{Hou:2020rse}
Jiamin Hou et~al.
\newblock {The Completed SDSS-IV extended Baryon Oscillation Spectroscopic
  Survey: BAO and RSD measurements from anisotropic clustering analysis of the
  Quasar Sample in configuration space between redshift 0.8 and 2.2}.
\newblock {\em Mon. Not. Roy. Astron. Soc.}, 500(1):1201--1221, 2020.

\bibitem{Hwang:2022hla}
Seung-gyu Hwang, Benjamin L'Huillier, Ryan~E. Keeley, M.~James Jee, and Arman
  Shafieloo.
\newblock {How to use GP: Effects of the mean function and hyperparameter
  selection on Gaussian Process regression}.
\newblock 6 2022.

\bibitem{Jimenez:2001gg}
Raul Jimenez and Abraham Loeb.
\newblock {Constraining cosmological parameters based on relative galaxy ages}.
\newblock {\em Astrophys. J.}, 573:37--42, 2002.

\bibitem{Joyce:2016vqv}
Austin Joyce, Lucas Lombriser, and Fabian Schmidt.
\newblock {Dark Energy Versus Modified Gravity}.
\newblock {\em Ann. Rev. Nucl. Part. Sci.}, 66:95--122, 2016.

\bibitem{Keeley:2020aym}
Ryan~E. Keeley, Arman Shafieloo, Gong-Bo Zhao, Jose~Alberto Vazquez, and
  Hanwool Koo.
\newblock {Reconstructing the Universe: Testing the Mutual Consistency of the
  Pantheon and SDSS/eBOSS BAO Data Sets with Gaussian Processes}.
\newblock {\em Astron. J.}, 161(3):151, 2021.

\bibitem{Khadka:2021vqa}
Narayan Khadka, Orlando Luongo, Marco Muccino, and Bharat Ratra.
\newblock {Do gamma-ray burst measurements provide a useful test of
  cosmological models?}
\newblock {\em JCAP}, 09:042, 2021.

\bibitem{Koyama:2015vza}
Kazuya Koyama.
\newblock {Cosmological Tests of Modified Gravity}.
\newblock {\em Rept. Prog. Phys.}, 79(4):046902, 2016.

\bibitem{Krishnan:2021dyb}
Chethan Krishnan, Roya Mohayaee, Eoin~\'O. Colg\'ain, M.~M. Sheikh-Jabbari, and
  Lu~Yin.
\newblock {Does Hubble tension signal a breakdown in FLRW cosmology?}
\newblock {\em Class. Quant. Grav.}, 38(18):184001, 2021.

\bibitem{LHuillier:2018rsv}
Benjamin L'Huillier, Arman Shafieloo, Eric~V. Linder, and Alex~G. Kim.
\newblock {Model Independent Expansion History from Supernovae: Cosmology
  versus Systematics}.
\newblock {\em Mon. Not. Roy. Astron. Soc.}, 485(2):2783--2790, 2019.

\bibitem{Linden2009CosmologicalPE}
Sebastian Linden, J.~M. Virey, and Andr'e Tilquin.
\newblock Cosmological parameter extraction and biases from type ia supernova
  magnitude evolution.
\newblock {\em Astronomy and Astrophysics}, 506:1095--1105, 2009.

\bibitem{Linder:2002et}
Eric~V. Linder.
\newblock {Exploring the expansion history of the universe}.
\newblock {\em Phys. Rev. Lett.}, 90:091301, 2003.

\bibitem{Liu:2020pfa}
Yuting Liu, Shuo Cao, Tonghua Liu, Xiaolei Li, Shuaibo Geng, Yujie Lian, and
  Wuzheng Guo.
\newblock {Model-independent constraints on cosmic curvature: implication from
  updated Hubble diagram of high-redshift standard candles}.
\newblock {\em Astrophys. J.}, 901(2):129, 2020.

\bibitem{Lonappan:2017lzt}
Anto~I. Lonappan, Sumit Kumar, Ruchika, Bikash~R. Dinda, and Anjan~A. Sen.
\newblock {Bayesian evidences for dark energy models in light of current
  observational data}.
\newblock {\em Phys. Rev. D}, 97(4):043524, 2018.

\bibitem{Malquarti:2003hn}
Michael Malquarti, Edmund~J. Copeland, and Andrew~R. Liddle.
\newblock {K-essence and the coincidence problem}.
\newblock {\em Phys. Rev. D}, 68:023512, 2003.

\bibitem{Moresco:2022phi}
Michele Moresco et~al.
\newblock {Unveiling the Universe with Emerging Cosmological Probes}.
\newblock 1 2022.

\bibitem{Peebles:2002gy}
P.~J.~E. Peebles and Bharat Ratra.
\newblock {The Cosmological Constant and Dark Energy}.
\newblock {\em Rev. Mod. Phys.}, 75:559--606, 2003.

\bibitem{Perivolaropoulos:2021jda}
Leandros Perivolaropoulos and Foteini Skara.
\newblock {Challenges for \ensuremath{<}math altimg=''si238.svg''
  display=''inline'' id=''d1e11032''\ensuremath{>}\ensuremath{<}mi
  mathvariant=''normal''\ensuremath{>}\ensuremath{\Lambda}\ensuremath{<}/mi\ensuremath{>}\ensuremath{<}/math\ensuremath{>}CDM:
  An update}.
\newblock {\em New Astron. Rev.}, 95:101659, 2022.

\bibitem{SupernovaCosmologyProject:1997zqe}
S.~Perlmutter et~al.
\newblock {Discovery of a supernova explosion at half the age of the Universe
  and its cosmological implications}.
\newblock {\em Nature}, 391:51--54, 1998.

\bibitem{SupernovaCosmologyProject:1998vns}
S.~Perlmutter et~al.
\newblock {Measurements of $\Omega$ and $\Lambda$ from 42 high redshift
  supernovae}.
\newblock {\em Astrophys. J.}, 517:565--586, 1999.

\bibitem{Pinho:2018unz}
Ana~Marta Pinho, Santiago Casas, and Luca Amendola.
\newblock {Model-independent reconstruction of the linear anisotropic stress
  $\eta$}.
\newblock {\em JCAP}, 11:027, 2018.

\bibitem{GpRasWil}
Carl~Edward Rasmussen and Christopher K.~I. Williams.
\newblock {\em Gaussian Processes for Machine Learning}.
\newblock The MIT Press, second edition, 2006.

\bibitem{Riess:2020fzl}
Adam~G. Riess, Stefano Casertano, Wenlong Yuan, J.~Bradley Bowers, Lucas Macri,
  Joel~C. Zinn, and Dan Scolnic.
\newblock {Cosmic Distances Calibrated to 1\% Precision with Gaia EDR3
  Parallaxes and Hubble Space Telescope Photometry of 75 Milky Way Cepheids
  Confirm Tension with $\Lambda$CDM}.
\newblock {\em Astrophys. J. Lett.}, 908(1):L6, 2021.

\bibitem{SupernovaSearchTeam:1998fmf}
Adam~G. Riess et~al.
\newblock {Observational evidence from supernovae for an accelerating universe
  and a cosmological constant}.
\newblock {\em Astron. J.}, 116:1009--1038, 1998.

\bibitem{Sahni:1999gb}
Varun Sahni and Alexei~A. Starobinsky.
\newblock {The Case for a positive cosmological Lambda term}.
\newblock {\em Int. J. Mod. Phys. D}, 9:373--444, 2000.

\bibitem{Pan-STARRS1:2017jku}
D.~M. Scolnic et~al.
\newblock {The Complete Light-curve Sample of Spectroscopically Confirmed SNe
  Ia from Pan-STARRS1 and Cosmological Constraints from the Combined Pantheon
  Sample}.
\newblock {\em Astrophys. J.}, 859(2):101, 2018.

\bibitem{Seikel_2012}
Marina Seikel, Chris Clarkson, and Mathew Smith.
\newblock Reconstruction of dark energy and expansion dynamics using gaussian
  processes.
\newblock {\em Journal of Cosmology and Astroparticle Physics},
  2012(06):036--036, jun 2012.

\bibitem{Shafieloo_2012}
Arman Shafieloo, Alex~G. Kim, and Eric~V. Linder.
\newblock Gaussian process cosmography.
\newblock {\em Physical Review D}, 85(12), jun 2012.

\bibitem{Thakur:2012rp}
Shruti Thakur, Akhilesh Nautiyal, Anjan~A Sen, and T~R Seshadri.
\newblock {Thawing Versus. Tracker Behaviour: Observational Evidence}.
\newblock {\em Mon. Not. Roy. Astron. Soc.}, 427:988--993, 2012.

\bibitem{Tsujikawa:2010zza}
Shinji Tsujikawa.
\newblock {Modified gravity models of dark energy}.
\newblock {\em Lect. Notes Phys.}, 800:99--145, 2010.

\bibitem{Tutusaus:2018ulu}
Isaac Tutusaus, Brahim Lamine, and Alain Blanchard.
\newblock {Model-independent cosmic acceleration and redshift-dependent
  intrinsic luminosity in type-Ia supernovae}.
\newblock {\em Astron. Astrophys.}, 625:A15, 2019.

\bibitem{Vagnozzi:2019ezj}
Sunny Vagnozzi.
\newblock {New physics in light of the $H_0$ tension: An alternative view}.
\newblock {\em Phys. Rev. D}, 102(2):023518, 2020.

\bibitem{Velten:2014nra}
H.E.S. Velten, R.F. vom Marttens, and W.~Zimdahl.
\newblock {Aspects of the cosmological \textquotedblleft{}coincidence
  problem\textquotedblright{}}.
\newblock {\em Eur. Phys. J. C}, 74(11):3160, 2014.

\bibitem{Wang:2019yob}
Bo~Wang, Jing-Zhao Qi, Jing-Fei Zhang, and Xin Zhang.
\newblock {Cosmological Model-independent Constraints on Spatial Curvature from
  Strong Gravitational Lensing and SN Ia Observations}.
\newblock {\em Astrophys. J.}, 898(2):100, 2020.

\bibitem{Wei:2008kq}
Hao Wei and Shuang~Nan Zhang.
\newblock {Reconstructing the cosmic expansion history up to redshift z=6.29
  with the calibrated gamma-ray bursts}.
\newblock {\em Eur. Phys. J. C}, 63:139--147, 2009.

\bibitem{Weinberg_2013}
David~H. Weinberg, Michael~J. Mortonson, Daniel~J. Eisenstein, Christopher
  Hirata, Adam~G. Riess, and Eduardo Rozo.
\newblock Observational probes of cosmic acceleration.
\newblock {\em Physics Reports}, 530(2):87--255, sep 2013.

\bibitem{williams1995gaussian}
Christopher Williams and Carl Rasmussen.
\newblock Gaussian processes for regression.
\newblock {\em Advances in neural information processing systems}, 8, 1995.

\bibitem{2011NatPh...7Q.833W}
Alison {Wright}.
\newblock {Nobel Prize 2011: Perlmutter, Schmidt \& Riess}.
\newblock {\em Nature Physics}, 7(11):833, November 2011.

\bibitem{Yoo:2012ug}
Jaewon Yoo and Yuki Watanabe.
\newblock {Theoretical Models of Dark Energy}.
\newblock {\em Int. J. Mod. Phys. D}, 21:1230002, 2012.

\bibitem{Zheng:2020tau}
Xiaogang Zheng, Shuo Cao, Yuting Liu, Marek Biesiada, Tonghua Liu, Shuaibo
  Geng, Yujie Lian, and Wuzheng Guo.
\newblock {Model-independent constraints on cosmic curvature: implication from
  the future space gravitational-wave antenna DECIGO}.
\newblock {\em Eur. Phys. J. C}, 81(1):14, 2021.

\bibitem{Zlatev:1998tr}
Ivaylo Zlatev, Li-Min Wang, and Paul~J. Steinhardt.
\newblock {Quintessence, cosmic coincidence, and the cosmological constant}.
\newblock {\em Phys. Rev. Lett.}, 82:896--899, 1999.

\end{thebibliography}






\end{document}